\documentclass[11pt]{article}
\newcommand{\MSbar}{\overline{\mbox{MS}}} 

\def\semi{;\hfil\break}
\def\npb{Nucl.\ Phys.\ }

\def\prl{Phys.\ Rev.\ Lett.\ }

\def\Tr{\hbox{Tr}}

\def\DRED{\ifmmode{{\rm DRED}} \else{{DRED}} \fi}
\newcommand{\bfn}{$\beta$-function}
\newcommand{\scl}[1]{\ensuremath{\phi^{#1}}}
\newcommand{\sclrep}[1]{\ensuremath{S^{#1}}}
\newcommand{\fermrep}[1]{\ensuremath{R^{#1}}}
\newcommand{\gaggrp}{\ensuremath{G}}
\newcommand{\beq}{\begin{eqnarray}}
\newcommand{\eeq}{\end{eqnarray}}
\newcommand{\eqlab}[1]{\label{eqn:#1}}

\newcommand{\bibref}[1]{\cite{#1}}
\newcommand{\fulleqref}[1]{Eq.~\eqref{#1}}
\newcommand{\eqref}[1]{(\ref{eqn:#1})}

\newcommand{\dreg}{DREG}
\newcommand{\dred}{DRED}
\newcommand{\msbar}{\ensuremath{\overline{\mbox{MS}}}}
\newcommand{\gbare}{\ensuremath{g_0}}
\newcommand{\Form}{{\scshape Form}}
\newcommand{\Mincer}{ {\scshape Mincer}}
\newcommand{\qgraf}{ {\scshape Qgraf}}
\newcommand{\Mathematica}{{\scshape Mathematica}}
\newcommand{\Feynalyse}{{\scshape Feynalyse}}
\newcommand{\btwid}[1]{\ensuremath{\tilde{\beta_{#1}}}}
\newcommand{\pole}[1]{\ensuremath{\mathrm{Pole}\left[#1\right]}}

\begin{document}
\hfill IUHET 434

\hfill LTH 498  

\hfill April 2001

{\begin{center} 
{\LARGE {Three loop gauge $\beta$-function for the most general 
single gauge-coupling theory} } \\ [8mm] 
{\large A.G.M. Pickering$^{a,b}$, J.A. Gracey$^b$ \& D.R.T. Jones$^b$ \\ 
[3mm] }
\end{center} 
} 

\vspace{2cm} 
\begin{itemize} 
\item[$^a$] Physics Department, Indiana University, Bloomington, IN 47405, USA. 
\item[$^b$] Theoretical Physics Division, Department of Mathematical
Sciences, \\ 
University of Liverpool, Peach Street, Liverpool, L69 7ZF, United Kingdom. 
\end{itemize} 

\vspace{5cm} 
\noindent 
{\bf Abstract.} 
We calculate the three loop contribution to the $\beta$-function of the gauge
coupling constant in a general, anomaly-free, renormalisable gauge field theory
involving a single gauge coupling using the background field method in the 
$\MSbar$ scheme. 

\newpage 

The renormalisation group functions play an important role in governing the 
fundamental properties of quantum field theories. For example asymptotic 
freedom is a consequence of the sign of the one loop contribution to the gauge
$\beta$-function, $\beta_g$,  for a non-abelian gauge theory~\cite{GRP}. The 
variations with scale of the coupling constants of a given theory are 
established from knowledge of the $\beta$-functions, and so to relate physics 
at different scales in a given theory it is important to know the 
$\beta$-functions as accurately as possible. For QCD, there has been impressive
progress recently with the provision of the four loop term for $\beta_g$ in the 
$\MSbar$ scheme,\cite{RVL}. This  extended the scheme independent two loop 
results of~\cite{CJ} and the three loop calculation of~\cite{TVZ,LV}. Evidently
QCD does not, however, represent the most general renormalisable anomaly-free 
gauge field theory which one could consider in four dimensions; such a theory 
would involve both fermions and scalars, with Yukawa and quartic scalar
interactions. Clearly this general gauge theory would contain the standard 
model (SM) as a particular case. The one loop $\beta$-functions for this most 
general theory, were first written down in~\cite{CEL}, whilst the two loop 
corrections were calculated in the $\MSbar$ scheme in~\cite{JO,MV1}. 

Although the result of~\cite{RVL} represents substantial progress in refining 
the running of the strong coupling constant, it transpires that the 
renormalisation of the most general gauge theory has not yet been performed at 
{\em three} loops. In this letter, therefore, we extend the result 
of~\cite{TVZ,LV} for $\beta_g^{(3)}$ to include Yukawa and quartic scalar 
interactions. We perform the calculation in such a way that the expressions for
both the conventional dimensional regularisation MS (or $\MSbar$) scheme and 
(in the supersymmetric case) the dimensional reduction, (DRED), 
scheme~\cite{SIEG} can be easily extracted. In the latter case we can compare 
our result with that of~\cite{JJN}. This is especially worthwhile since the
result of~\cite{JJN} was arrived at indirectly in the non-abelian case.
Therefore, in fact, our results will include the first {\it explicit\/} 
calculation of $\beta_g^{(3)\DRED}$. We defer to a later 
publication~\cite{PGJ}, the full details of the calculation where we will also 
provide the extension to the multi-gauge coupling case, the three loop $\MSbar$ 
expressions of the remaining renormalisation group functions of the other 
couplings and the results in the special case of the SM.

Another motivation for undertaking this three loop renormalisation rests in
recent exciting developments for constructing four dimensional conformal field 
theories, which are not necessarily supersymmetric, based on ideas from the 
Maldacena conjecture~\cite{MAL}. In~\cite{KS}, Kachru and Silverstein 
constructed four dimensional field theories on orbifolds so that (at least in 
the large $N_{\!c}$ limit) their $\beta$-functions vanished, leading to 
(approximate) conformal symmetry. Inspired by this, Lawrence et al~\cite{LNV} 
conjectured that the orbifold approach could be refined so that the field 
theories were conformal for {\em finite} $N_{\!c}$ with and without 
supersymmetry. Moreover, it was proposed in~\cite{FV} that this framework could
provide a non-supersymmetric resolution of the gauge hierarchy problem. In 
essence, above a certain energy scale of the order of $1-10$ TeV, a conformal 
field theory exists on an orbifold. The low energy theory (obtained by breaking
the conformal invariance softly, by the addition of operators of dimension less
than four) can be phenomenologically viable~\cite{F1}. 
Subsequently~\cite{FV}-\cite{CST}, the abelian and non-abelian orbifold models 
satisfying these criteria were constructed. The method used was to impose that 
all the renormalisation group functions of the most general four dimensional 
gauge theory vanish, thereby ensuring conformality. For instance, the results 
for $\beta_g^{(2)}$ and the one loop scalar and Yukawa $\beta$-functions have 
been analysed. This is a systematic approach since it determines precisely the 
(non-supersymmetric) models which are not just conformal at large $N_{\!c}$ but
also when $N_{\!c}$ is finite. A large set of possible theories emerged; 
however it is as yet unclear as to which is the strongest candidate for 
realistic physics beyond the SM. Moreover, it may be the case that the higher 
order corrections to the renormalisation group functions could restrict this 
set of such possible models further~\cite{F3,CST}. Thus our provision of new 
results at three loops here is relevant.

We consider a general renormalisable quantum field theory defined by the 
following Lagrangian, $\cal L$:
\beq 
{\cal L } &=& -\frac{1}{4} ( G_{\mu\nu}^A )^2 \,+ \,  i\, \psi_j   
\sigma^\mu D_\mu \bar{\psi}^j \,+\,\frac{1}{2} \,|D_\mu \phi |^2 \,- 
\,\frac{1}{2} \,(Y^{aij}\phi^a \psi_i\psi_j \,+\, 
\bar{Y}^{a}_{ij}\phi^a \bar{\psi}^i\bar{\psi}^j)  \nonumber \\
& &  - \,\,\frac{1}{4!} \lambda_{abcd} \phi^a\phi^b\phi^c\phi^d \,\,+ \,\, 
\mathrm{gauge \,\, fixing \,\,} \,\,+ \,\, \mathrm{ghosts} \,\,+ \,\, 
\mathrm{mass \,\,\, terms} 
\eqlab{lagdef}
\eeq 
where $\lambda_{abcd}$ is totally symmetric, $Y^{aij}$ and its complex 
conjugate $\bar{Y}^{a}_{ij}$ are symmetric on their $ij$ indices and $D_\mu$ is
the covariant derivative, $\partial_\mu - i g A_\mu$.

We extract divergences using dimensional regularisation, \dreg, with modified 
minimal subtraction, \msbar, in $d = 4-2\epsilon$ dimensions. The mass terms in
the Lagrangian have been suppressed because they do not contribute to $\beta_g$
in a mass-independent renormalisation scheme, such as \msbar. Gauge fixing and 
ghosts will be discussed below. 

Without loss of generality the scalar fields, \scl{a}, are taken to be real. As
a result they transform in an antisymmetric representation, \sclrep{A}, of the 
Lie algebra of the gauge group, \gaggrp, where 
\beq
\sclrep{A}_{ab} \,\,=\,\, - \sclrep{A}_{ba}, \,\,\,\,\,\,\,\,\,\,
\left[ \sclrep{A}, \sclrep{B} \right] \,\,=\,\, i f^{ABC} \sclrep{C},
\eqlab{sclalg}
\eeq
and $f^{ABC}$ are the structure constants of \gaggrp. 

We work with two-component fermions, $\psi_j$, and their conjugates, 
$\bar{\psi}^j$, so chiral theories can be considered. The fermions transform in
a representation, \fermrep{A}, of the Lie algebra of \gaggrp, where 
\beq
\fermrep{A} \,\,=\,\, \fermrep{A\dagger}, \,\,\,\,\,\,\,\,\,\,
\left[ \fermrep{A}, \fermrep{B} \right] \,\,=\,\, i f^{ABC} \fermrep{C}.
\eqlab{fermalg}
\eeq
For cancellation of gauge anomalies we require
\beq
\mathrm{\Tr}[R^A \{ R^B, R^C \} ] \,\, = \,\, 0.
\eqlab{anomc}
\eeq

In this paper, we restrict our attention to simple gauge groups and hence a 
single gauge coupling constant, $g$ (of course the case of QED can easily be 
deduced as well). The extension to the case of several gauge couplings is 
non-trivial (unlike at one and two loops), and will be presented 
elsewhere~\bibref{PGJ} along with explicit results for the SM. 

The \bfn\ for the renormalised coupling, $g$, is defined as:
\beq
\beta_g \,\,=\,\, \mu \frac{d g}{d\mu},
\,\,\,\,\,\,\mathrm{with}\,\,\,\,\,\, 
\gbare \,\, = \mu^\epsilon Z_g g
\eqlab{gbaredef}
\eeq
where \gbare\ is the bare coupling constant, and $\mu$ is the renormalisation 
scale. To find $\beta_g$, we must first calculate $Z_g$. One might consider 
calculating, for example, the $3$-point gauge-particle vertex. The graphs which
contribute to this can be generated by attaching three external gauge legs in 
all possible ways to vacuum graphs; at three loops this gives rise to an 
enormous number of graphs. In addition one would of course have to calculate 
the gauge-particle renormalization factor, $Z_A$. Evaluating this to three 
loops involves another significant calculation, this  time of the $2$-point 
gauge self-energy. Moreover, difficulties arise when we consider performing the
$3$-point momentum integrals required to three loop order. We will be using the
integration package \Mincer~\bibref{mincer} as implemented in the symbolic 
manipulation language \Form~\cite{form} and described in~\cite{mincerform}, 
which has the ability to handle only one independent external momentum and 
therefore we must set the  momentum on all but two of the external legs to 
zero. For graphs with more than two external legs, this leads to the potential 
introduction of spurious infrared divergences, which also show up as poles in 
$\epsilon$ in \dreg\ and are therefore difficult to distinguish from 
ultra-violet infinities. Whilst it is no doubt possible to perform the 
calculation this way, significant simplifications result from the use of the
background field method pioneered by DeWitt in~\cite{bfm} and later extended 
by 't Hooft~\cite{tHbfm}, DeWitt~\cite{DWbfm}, Boulware~\cite{Bbfm} and 
Abbott~\cite{Abbott:1981hw}, who performed a two loop pure Yang-Mills 
calculation in the Landau gauge ($\alpha = 0$), later generalised to 
$\alpha\neq 0$ by  Capper and Maclean~\cite{Capper:1982tf}. More recently the 
method has been used to study the renormalisation of gauge theories with a 
non-semisimple colour group,\cite{Grassi}. 

The background field method consists of splitting the gauge field, $A_\mu$ into
a quantum field, $Q_\mu$, and a background field, $B_\mu$, such that 
\beq 
A_\mu &=& Q_\mu + B_\mu. 
\eeq 
Only $Q_\mu$ is integrated over in the path integral and we a choose a 
gauge-fixing term for $Q_\mu$ which is invariant under gauge transformations 
with respect to $B_\mu$; the ghost vertices may then be derived from this in 
the usual way. Corrections to the effective action are gauge invariant with 
respect to $B_\mu$ and as a result it can be shown that the following relations
hold  
\beq Z_\alpha &=& Z_Q \nonumber\\ 
\sqrt{Z_B} Z_g &=& 1 
\eqlab{bfmrels}  
\eeq 
Thus $Z_g$, which we need to obtain $\beta_g$, may be obtained from the 
background gauge-field renormalisation constant, $Z_B$, which is easier to 
compute. Similarly we can deduce the gauge parameter renormalisation constant, 
$Z_\alpha$, from the quantum field renormalisation constant, $Z_Q$, and, as 
discussed below, this enables us to check our results by performing the 
alculation for arbitrary $\alpha$. Consequently, both $Z_g$ and $Z_\alpha$ can 
be found from calculations which only involve evaluating $2$-point graphs. 
There are clearly significantly fewer of these graphs and they are more easily
evaluated  using \Mincer\ as there is no need to nullify any external momenta. 

In \msbar, $Z_g$ and hence $Z_B$ may be written as a sum of poles in 
$\epsilon$. We define  
\beq
Z_B &=& 1 + \sum_{n=1}^{\infty} \frac{\Delta_n}{\epsilon^n},
\eeq
and it follows from Eqs.~\eqref{gbaredef} and \eqref{bfmrels} that 
\beq
g_0^2 = \mu^{2\epsilon}Z_B^{-1}g^2.
\eqlab{gsquare}
\eeq
$\beta_g$ is determined by $\Delta_1$, the simple pole in $Z_B$; and at any 
given order in perturbation theory, the higher order poles $\Delta_n$ can be 
calculated from the simple pole at lower orders. We have:
\beq
\btwid{g} = \,\,-  \frac{g}{2} \left( g \frac{\partial }{\partial g} + Y . 
\frac{\partial }{\partial Y} + \bar{Y} . \frac{\partial }{\partial
\bar{Y}}  +
2 \lambda . \frac{\partial }{\partial \lambda} \right)\Delta_{1}
\eqlab{betadef}
\eeq
and
\beq
\btwid{g} \, \Delta_n +  \frac{g}{2} \left( g \frac{\partial }{\partial g}
+ Y.\frac{\partial }{\partial Y} + \bar{Y} . \frac{\partial }{\partial
\bar{Y}} + 2 \lambda . \frac{\partial }{\partial \lambda} \right)\Delta_{n+1}
\nonumber \\ 
= \,\, \frac{g}{2} \left( \btwid{g} \frac{\partial }{\partial g} + 
 \tilde{\beta}_Y . \frac{\partial }{\partial Y} + 
 \tilde{\beta}_{\bar{Y}} . \frac{\partial }{\partial \bar{Y}} +  
\tilde{\beta}_\lambda . \frac{\partial }{\partial \lambda} \right)
\Delta_n 
\eqlab{betapoles}
\eeq
where the reduced \bfn s, $\tilde{\beta}$ are independent of $\epsilon$ and 
\beq
\btwid{g} \,\,=\,\, \beta_g + \epsilon g, \,\,\,\,\, 
\btwid{Y} \,\,=\,\, \beta_Y + \epsilon Y, \,\,\,\,\, 
\btwid{\lambda} \,\,=\,\, \beta_\lambda + 2 \,\epsilon \lambda.  
\eqlab{defredbeta}
\eeq
\fulleqref{betapoles} can be used to verify that the higher order poles 
generated in the final result for $\beta_g$ are consistent with the lower
order results thus providing a strong check on any calculation. 

The first step in the calculation of $Z_B$ is to generate all of the relevant 
graphs along with their appropriate combinatoric factors. We found it useful to
write a package in \Mathematica~\cite{mathem}, called \Feynalyse, which, among 
other things, is capable of generating all of the required graphs starting from
the Lagrangian, $\cal L$. (As a check, we also generated the graphs using 
\qgraf~\cite{qgraf}.) By producing the graphs in a \Mathematica\ environment we
are then able to manipulate them using other routines in \Feynalyse, which can 
draw and label each graph systematically, calculate combinatoric factors and
signs due to fermion loops, assign Feynman rules to the vertices and
propagators and use these to generate the necessary mathematical expressions 
for each graph. Once the complete set of graphs, signs and combinatoric factors
has been found, the remaining calculation can be factorised into two 
independent calculations: the  group theoretical factor and the momentum 
integration. The only vertex for which this causes a problem is the $4$-point 
gauge-particle interaction  discussed in more detail in~\cite{PGJ}. 

The group theory is dealt with in \Mathematica\ by using a set of specially 
written transformation rules which allow any group theory expression up to 
three loops to be reduced to a sum of linearly independent terms. It is 
important to construct such a set to ensure that all necessary cancellations 
between similar terms are carried out. We make use of the Lie algebra of the 
group generators, \sclrep{A} and \fermrep{A}, defined in Eqs.~\eqref{sclalg} 
and \eqref{fermalg}, the Jacobi relation for the structure constants, $f^{ABC}$
and the following relations 
\beq
{R^{Aj}_{k}} Y^{aik} + {R^{Ai}_{k}} Y^{akj} + Y^{bij} S^A_{ba} = 0, 
\\ 
\lambda_{ebcd} S^A_{ea} + \lambda_{aecd} S^A_{eb} + \lambda_{abed}
S^A_{ec} + 
\lambda_{abce} S^A_{ed} = 0,
\eqlab{grpthy} 
\eeq
which follow from gauge invariance. Most of the group theoretical factors we 
encounter up to three loops can be expressed in terms of the quadratic Casimirs
of the relevant representations of the Lie algebra of \gaggrp. These are
defined as follows:
\beq
\Tr(\sclrep{A}\sclrep{B}) \,\,=\,\, \delta^{AB} \,T(S) & &
\sclrep{A}_{ac} \sclrep{A}_{cb} \,\,=\,\, C(S)_{ab} \\
\Tr(\fermrep{A}\fermrep{B}) \,\,=\,\, \delta^{AB} \,T(R) & &
R^{Ak}_{i} R_k^{Aj} \,\,=\,\, {C(R)_{i}}^j \\
f^{ACD} f^{BCD} \,\,=\,\, \delta^{AB} \,C(G) & & \delta^{AA} \,\,=\,\, r . 
\eeq
As mentioned above, we employ the anomaly cancellation condition, 
\fulleqref{anomc}. This justifies our use of the na{\"{\i}}ve $\gamma_5$ matrix
in defining two component spinors, because, at three loops, graphs containing 
two one loop fermion triangles are the only graphs which can generate 
non-trivial corrections to the gauge propagator involving the  
$\epsilon^{\mu\nu\rho\sigma}$ tensor. As explained in detail in~\cite{PGJ}, 
\fulleqref{anomc} means that we can set $\epsilon^{\mu\nu\rho\sigma}$ to zero, 
thereby avoiding inconsistencies with the $\gamma$-algebra in $d$-dimensions.

The momentum integrations are performed using \Mincer~\cite{mincer,mincerform}.
We use it to  calculate the divergent part of each diagram, expressed as poles 
in $\epsilon$. Although we need only calculate the simple  pole to deduce 
$\beta_g$, we keep higher order poles as a check on our results. By retaining 
all the poles and the appropriate finite terms for each diagram at one, two and
three loops, we can perform the subtraction of subdivergences in one step, at 
the end of the calculation, as explained below. First, though, we must 
calculate the unsubtracted divergences for all graphs contributing to the gauge
propagator up to three loops. In order to do this using \Mincer\ it is 
essential to label the momenta in the graph according to its `topology', 
exactly as defined in the \Mincer\ documentation,\cite{mincer}. Each `topology'
corresponds to a particular integration routine which assumes that the momenta 
are already correctly labelled when it is called. Due to the large number of 
graphs it is necessary to construct a program which can work out which 
integration package needs to be called to integrate a given graph and label its
momenta accordingly. The program we have written uses the definition of the 
graph in \Mathematica, produced by our package \Feynalyse, and the 
corresponding Feynman rules to produce a complete \Form\ file containing the 
necessary scalar integrals along with all header files and definitions needed 
by \Form\ and \Mincer, as well as the correct call to the appropriate 
integration package for that topology. The file is completely self contained 
and can be read by \Form\ which performs the integrations using \Mincer\ and 
stores the results on disk. These are then collected and read into 
\Mathematica\ by a further set of routines in \Feynalyse, which combine the 
results with the simplified group theory factors and produce a result for the
total contribution to the effective action from all graphs. By following this 
procedure we can evaluate the unsubtracted contributions to the gauge 
propagator at one, two and three loops as a function of the bare couplings. 
However, as we are calculating a three loop result we must ensure that the 
Lagrangian has been renormalised up to two loops. This means that there are 
additional one and two loop counterterms which, when inserted in graphs with 
one and two loop topologies, give rise to contributions at the three loop 
level. By multiplicative renormalisability, the counterterms have the same form
as the tree level interactions and to calculate these contributions 
independently would be extremely inefficient. Therefore, we adopt the procedure
used in~\cite{LV} for the subtraction of subdivergences, which is summarised 
below. 

We begin by introducing a loop-counting parameter $h$, by replacing  
\beq
g_0 \,\,\rightarrow\,\, h \, g_0, \,\,\,\,\,\,\, 
Y_0 \,\,\rightarrow\,\, h \, Y_0, \,\,\,\,\,\,\, 
\lambda_0 \,\,\rightarrow\,\, h^2 \,\lambda_0. 
\eeq 
The corrections to the $2$-point gauge-field interaction must be transverse, so
that:
\beq
\Gamma_{0\mu\nu}^{BB} &=& \Gamma_{0}^{BB} \left( g_{\mu\nu} 
- \frac{q_\mu q_\nu}{q^2} \right),
\eeq
where $q$ is the external momentum, and 
\beq
\Gamma^{BB} &=& Z_B \Gamma_{0}^{BB}.
\eqlab{gamdef}
\eeq
Here $\Gamma^{BB}$ is finite (as $\epsilon\to 0$) when expressed in terms of 
the renormalised couplings. We have calculated 
\beq
\Gamma^{BB}_0 &=& 1 + A_0\, h^2 + B_0\,h^4 + C_0\, h^6 + \ldots
\eqlab{finite}
\eeq 
where $A_0, B_0,C_0\cdots$ are functions of the bare couplings $g_0$, $Y_0$, 
$\lambda_0$, $\alpha_0$. First we expand $Z_B$ in loops so that   
\beq
Z_B \,\,\, = \,\,\, 1 + \sum_{n=1}^{\infty} \frac{\Delta_n}{\epsilon^n}
&=& 
1 + a h^2 + b h^4 + c h^6 + \ldots 
\eqlab{Zdef}
\eeq
Then we re-express $\Gamma^{BB}_0$ in terms of renormalised couplings using 
\fulleqref{gsquare} and similarly for the other bare couplings. (In fact we do 
not need to renormalise $\lambda_{0abcd}$ as it appears for the first time at 
three loops and hence we can replace it with the renormalised value, 
$\lambda_{abcd}$. We will however require $Z_\alpha$ at two loops and $Z_Y$, 
for the Yukawa couplings, at one loop.) Note that from \fulleqref{Zdef}, 
$a$, $b$, $c$ are functions of the renormalised couplings, $g$, $Y$, $\lambda$,
$\alpha$ and contain only pole terms in $\epsilon$. We thus obtain  
\beq
\Gamma^{BB}_0 &=& 1 + A\, h^2 + B\,h^4 + C\, h^6 + \ldots
\eqlab{finiteb}
\eeq
where $A,B,C\ldots$ are functions of the renormalised couplings. Now 
substituting Eqs.~\eqref{Zdef}, \eqref{finiteb}\ in Eq.~\eqref{gamdef}, we 
simply impose that $\Gamma^{BB}$ is finite as $\epsilon\to 0$ to deduce $Z_B$. 
We thus have: 
\beq
a &=& - \pole{A} \nonumber \\ 
b &=& - \pole{B + a A}  \\ 
c &=& - \pole{C + B a + A b},  \nonumber
\eeq
where $\pole{x}$ denotes the pole part of $x$ with respect to $\epsilon$. We 
stress that it is important to keep some of the finite corrections to the lower
order one and two loop graphs (ie $A$ and $B$) in order to calculate $a$, $b$
and $c$ correctly.  

We define the loop expansion of the \bfn\ to be 
\beq
\tilde{\beta}_g &=& \tilde{\beta}_g^{(1)} h^2 + \tilde{\beta}_g^{(2)} h^4
+ \tilde{\beta}_g^{(3)} h^6 + \ldots 
\eeq
Similarly, Eq. \eqref{Zdef} implies that 
\beq
\Delta_1 &=& a_1 h^2 + b_1 h^4 + c_1 h^6 + \ldots
\eeq
where we would expect that 
\beq
a = \frac{a_1}{\epsilon},\quad b = \frac{b_1}{\epsilon} +
\frac{b_2}{\epsilon^2}\quad\hbox{and}\quad
c = \frac{c_1}{\epsilon} + \frac{c_2}{\epsilon^2} +
\frac{c_3}{\epsilon^3},
\eeq
but it follows from \fulleqref{betapoles}\ that $b_2 = c_3 = 0$. Using 
\fulleqref{betadef}, we find that 
\beq
\tilde{\beta}_g^{(1)} &=& - \, g  a_1 \\
\tilde{\beta}_g^{(2)} &=& - \, 2 g  b_1 \\
\tilde{\beta}_g^{(3)} &=& - \, 3 g  c_1 
\eeq
Our calculations verify that $b_2 = c_3 = 0$ and we find   
\beq
(16\pi^2) a_1 &=&  \frac{11\,C(G)\,g^2}{3} - \frac{2\,g^2\,T(R)}{3} - 
\frac{g^2\,T(S)}{6} \eqlab{oneresult}\\
(16\pi^2)^2 b_1 &=& \frac{17\,{C(G)}^2\,g^4}{3} - 
   \frac{5\,C(G)\,g^4\,T(R)}{3} - 
   \frac{C(G)\,g^4\,T(S)}{6} - \nonumber \\ & & 
   \frac{g^4\,\Tr({C(S)}^2)}{r} + 
   \frac{g^2\,\Tr(C(R) \bar{Y}^{a} Y^{a})}
    {2\,r} - \frac{g^4\,\Tr({C(R)}^2)}{\,r} 
\eqlab{onetworesult}
\eeq
\beq
(16\pi^2)^3 c_1 &=&  \frac{2857\,{C(G)}^3\,g^6}{162} - 
   \frac{1415\,{C(G)}^2\,g^6\,T(R)}{162} + 
   \frac{79\,C(G)\,g^6\,{T(R)}^2}{162} - \nonumber \\ & &
   \frac{545\,{C(G)}^2\,g^6\,T(S)}{648\,} + 
   \frac{29\,C(G)\,g^6\,T(R)\,T(S)}{162\,} - 
   \frac{C(G)\,g^6\,{T(S)}^2}{324\,} - \nonumber \\ & &
   \frac{1129\,C(G)\,g^6\,\Tr({C(S)}^2)}{216\,\,r} + 
   \frac{25\,g^6\,T(R)\,\Tr({C(S)}^2)}{54\,\,r} + \nonumber \\ & &
   \frac{49\,g^6\,T(S)\,\Tr({C(S)}^2)}{216\,\,r}  -   
   \frac{3\,C(G)\,g^4\,C(S)_{ab}\,\Tr(Y^{a}\bar{Y}^{b})}{4\,r} + 
\nonumber \\ & &     
   \frac{7\,g^4\,C(S)^{2}_{ab}\,\Tr(Y^{a}\bar{Y}^{b})}{6\,r} +
   \frac{2\,C(G)\,g^4\,\Tr(C(R)\bar{Y}^{a}Y^{a})}{r} +  \nonumber \\ & &   
   \frac{8\,g^4\,C(S)_{ab}\,\Tr(C(R)\bar{Y}^{a}Y^{b})}{3\,r} + 
   \frac{5\,g^4\,\Tr({C(R)}^2\bar{Y}^{a}Y^{b})}{12\,r} + \nonumber \\ & & 
   \frac{g^4\,\Tr(C(R)\bar{Y}^{a}C(R)Y^{a})}{12\,r} - 
   \frac{g^2\,C(S)_{ab}\,\Tr(Y^{c}\bar{Y}^{c}Y^{a}\bar{Y}^{b})}{6\,r} + 
\nonumber \\ & & 
   \frac{g^2\,C(S)_{ab}\,\Tr(Y^{c}\bar{Y}^{a}Y^{c}\bar{Y}^{b})}{6\,r} + 
   \frac{g^2\,C(S)_{ab}\,\Tr(\bar{Y}^{c}Y^{a}\bar{Y}^{c}Y^{b})}{6\,r} - 
\nonumber \\ & &
   \frac{g^2\,\Tr(C(R)\bar{Y}^{a}Y^{a}\bar{Y}^{b}Y^{b})}{24\,r} - 
   \frac{g^2\,\Tr(C(R)\,\bar{Y}^{a}Y^{b}\bar{Y}^{a}\,Y^{b})}{2\,r} -
\nonumber \\ & &
   \frac{g^2\,\Tr(R^{A}\bar{Y}^{a}Y^{b}R^{A}\bar{Y}^{b}Y^{a})}{3\,r} - 
   \frac{7\,g^2\,\Tr(Y^{a}\bar{Y}^{b})\Tr(\bar{Y}^{b}Y^{a}C(R))}{12\,r} + 
\nonumber \\ & & 
   \frac{g^2\,\Tr(Y^{a}\bar{Y}^{b})\Tr(Y^{b}\bar{Y}^{c})C(S)_{ca}}{12\,r}
- 
   \frac{g^2\,\Tr(C(R)\,\bar{Y}^{a}Y^{b}\bar{Y}^{b}\,Y^{a})}{8\,r}  - 
\nonumber \\ & &
   \frac{205\,C(G)\,g^6\,\Tr({C(R)}^2)}{54\,r} + 
   \frac{11\,g^6\,T(R)\,\Tr({C(R)}^2)}{27\,r} + \nonumber \\ & &  
   \frac{23\,g^6\,T(S)\,\Tr({C(R)}^2)}{108\,r} + 
   \frac{g^6\,\Tr({C(R)}^3)}{3\,r}  - 
   \frac{29\,g^6\,\Tr({C(S)}^3)}{12\,\,r} - \nonumber \\ & &  
\frac{g^4\,\lambda_{abcd}\,(\sclrep{A}\sclrep{B})_{ab}\,
    (\sclrep{A}\sclrep{B})_{cd} }{6\,r}  + 
\frac{g^2\,C(S)_{ab}\lambda_{acde} \lambda_{bcde}}{72\,r} 
\eqlab{thrloopspole}
\eeq
where the quartic scalar couplings makes its first appearance in the final 
line~\cite{CUR}. We also find  
\beq
(16\pi^2)^3 c_2 &=&  
  -\,\frac{187\,{C(G)}^3\,g^6}{27} + \frac{89\,{C(G)}^2\,g^6\,T(R)}{27} - 
  \frac{10\,C(G)\,g^6\,{T(R)}^2}{27} + \nonumber \\ & &
  \frac{14\,{C(G)}^2\,g^6\,T(S)}{27} - 
  \frac{7\,C(G)\,g^6\,T(R)\,T(S)}{54} -
  \frac{C(G)\,g^6\,{T(S)}^2}{108} + \nonumber \\ & &   
  \frac{11\,C(G)\,g^6\,\Tr({C(S)}^2)}{9\,r} - 
  \frac{2\,g^6\,T(R)\,\Tr({C(S)}^2)}{9\,r} - \nonumber \\ & &
  \frac{g^4\,\Tr({C(R)}^2\,\bar{Y}^{a}\,Y^{a})}{2\,r} -  
  \frac{g^4\, \Tr(C(R)\,\bar{Y}^{a}\,C(R)\, Y^{a})} {2\,r} 
+ \nonumber \\ & &
  \frac{g^2\,\Tr(C(R)\,\bar{Y}^{a}\,Y^{a}\,\bar{Y}^{b}\,Y^{b})}{12\,r} +  
  \frac{g^2\,\Tr(C(R)\,\bar{Y}^{a}\,Y^{b}\,\bar{Y}^{a}\,Y^{b})}{3\,r}
+ \nonumber \\ & &
  \frac{g^2\,\Tr(C(R)\,\bar{Y}^{a}\,Y^{b}\,\bar{Y}^{b}\,Y^{a})}{12\,r}  - 
  \frac{g^6\,T(S)\,\Tr({C(R)}^2)}{18\,r} + \nonumber \\ & &
  \frac{g^2\,\Tr(Y^{a}\bar{Y}^{b})\, \Tr( \bar{Y}^{b} Y^{a} C(R) )}{6\,r}
- \frac{g^6\,T(S)\,\Tr({C(S)}^2)}{18\,r} + 
\nonumber \\ & & 
  \frac{11\,C(G)\,g^6\,\Tr({C(R)}^2)}{9\,r} - 
  \frac{2\,g^6\,T(R)\,\Tr({C(R)}^2)}{9\,r} ~. 
\eqlab{thrloopdpole}
\eeq
In any calculation it is important to check the internal consistency of the 
computations as well as ensure that they agree with any results in the 
literature. Given that our calculation involves a large degree of automatic 
computation it is vital that we apply all possible checks to Eqs. 
\eqref{thrloopspole} and \eqref{thrloopdpole}. First we consider internal 
consistency. We have verified that the gauge self-energy is transverse, in 
other words that the sum of all the relevant Feynman diagrams yields zero when 
contracted with the external propagator momentum. We have also checked that 
$\beta_g$ is independent of the gauge parameter, $\alpha$. Notice also that 
there are no $\zeta (3)$-terms in \fulleqref{thrloopspole}; this non-trivial
cancellation was to be expected, occurring as it does in the special case of 
QCD. Finally, we found $b_2 = c_3 = 0$, in accordance with 
\fulleqref{betapoles} and, by calculating $\beta_Y$ to one loop we have 
explicitly verified that $c_2$ is predicted to be exactly as given in 
\fulleqref{thrloopdpole} above. All this is evidence that we have performed the
subtraction of subdivergences correctly.

In addition to these internal consistency checks there are also several 
existing calculations of the \bfn\ up to and including some three loop 
calculations for specific cases. Many of these are concerned with 
supersymmetric theories, using DRED,  and so we cannot obtain them easily from 
the \dreg\ results given above, as this regularisation scheme violates 
supersymmetry. However, we in fact performed  the computations~\cite{PGJ}, in 
such a manner that both \dreg\ and \dred\ results are easily extracted. The 
$\gamma$-algebra was done in  $d = D - 2 C \epsilon$ dimensions and the 
momentum integrals in $d = 4 - 2\epsilon$ dimensions. Lack of space prevents us
from reproducing the general result here. The \dreg\ results (which we give) 
are obtained by choosing $D=4$ and $C=1$. Choosing instead $D=4$ and $C=0$ we 
can obtain the \dred\ result for a general $N=1$ supersymmetric theory in four
dimensions. (A full justification of this is provided in~\cite{PGJ}.) We need 
not reproduce this result here as it agrees entirely with that of~\cite{JJN}. 
As mentioned earlier, in~\cite{JJN} the non-abelian result was actually deduced
from the explicit calculation for the {\it abelian\/} case. Therefore we have 
performed the first direct calculation of $\beta_g^{(3)}$ for a  general $N=1$ 
supersymmetric non-abelian theory. Evidently this supersymmetric result
includes the special cases of $N=2$  and $N=4$ theories; but as an additional 
check we can obtain these by setting  $C=0$, with $D=6$ and $D = 10$ 
respectively, thereby reproducing the results of~\cite{susybetas} for these 
cases. (Of course in both cases $\beta_g$ vanishes beyond one loop). Moreover, 
for the $N=4$ theory, but  now using \dreg, we  obtain exactly the same 
non-zero result first obtained in~\cite{susy4a}. Apart from supersymmetric 
results, we have also verified that both the one and two loop \bfn s agree with
those already given in the literature~\cite{JO,MV1}. In addition, by setting
the Yukawa and scalar couplings  to zero, we have checked that our results are 
in exact agreement with the calculation of the QCD $\MSbar$ \bfn\ at three 
loops given in~\cite{TVZ,LV}, but not calculated using the background field 
method.  Therefore our results pass a stringent series of tests and are in 
agreement with all existing results in the literature.

Our main result, \fulleqref{thrloopspole}, involves a lot of terms compared 
with the one and two loop expressions, Eqs.~\eqref{oneresult} and 
\eqref{onetworesult}. This is, of course, dictated by the structure of the 
Feynman diagram topologies at three loops. For instance, one can have two 
distinct fermion loops giving rise to terms with two traces over the Yukawa 
indices. We have expressed the results as far as possible in terms of the usual
colour group Casimirs, $C(S)$, $C(R)$ and $C(G)$; but at three loops two terms 
remain in which it is not possible to combine two generators, $R^A$, $S^A$, 
using their group properties, with one such term depending on the Yukawa 
couplings. It is interesting that this does not occur in the supersymmetric 
case. This might be a complication for the classification of non-supersymmetric
conformal field theories~\cite{FV}-\cite{CST}. 

\vspace{0.4cm} 
\noindent 
{\bf Acknowledgements.} 

This work is supported in part by the United States Department of Energy under 
grant number DE-FG02-91ER40661,  by the Leverhulme Trust, and by {\sc PPARC.} 
The calculations were performed with use of the symbolic manipulation packages 
\Form, \cite{form}, and \Mathematica\, \cite{mathem}.  



\begin{thebibliography}{99}
\bibitem{GRP} D.J. Gross \& F.J. Wilczek, Phys. Rev. Lett. {\bf 30}
(1973), 1343\semi
H.D. Politzer, Phys. Rev. Lett. {\bf 30} (1973) 1346.
\bibitem{RVL} T. van Ritbergen, J.A.M. Vermaseren \& S.A. Larin, Phys. Lett.
{\bf B400} (1997) 379. 
\bibitem{CJ} D.R.T. Jones, Nucl. Phys. {\bf B75} (1974) 531\semi
W.E. Caswell, Phys. Rev. Lett. {\bf 33} (1974) 244. 
\bibitem{TVZ} O.V. Tarasov, A.A. Vladimirov \& A.Yu. Zharkov, Phys. Lett. 
{\bf B93} (1980) 429.
\bibitem{LV} S.A. Larin \& J.A.M. Vermaseren, Phys. Lett. {\bf B303} (1993) 
334.
\bibitem{CEL} T.P. Cheng, E. Eichten \& L.F. Li, Phys. Rev. {\bf D9} (1974)
2259. 
\bibitem{JO} I. Jack \& H. Osborn, J. Phys. {\bf A16} (1983) 1101;
Nucl. Phys. {\bf B249} (1985) 472.
\bibitem{MV1} M.E. Machacek \& M.T. Vaughn, Nucl. Phys. {\bf B222} (1983) 83;
Nucl. Phys. {\bf B236} (1984) 221;
Nucl. Phys. {\bf B249} (1985) 70.
\bibitem{SIEG} W. Siegel, Phys. Lett. {\bf B84} (1979) 193\semi  
D.M. Capper, D.R.T. Jones \& P. van Nieuwenhuizen, Nucl. Phys. {\bf B167} 
(1980) 479.
\bibitem{JJN} I. Jack, D.R.T. Jones \& C.G. North, Phys. Lett. {\bf B386} 
(1996) 138. 
\bibitem{PGJ} A.G.M. Pickering, J.A. Gracey \& D.R.T. Jones, paper in 
preparation.
\bibitem{MAL} J. Maldacena, Adv. Theor. Math. Phys. {\bf 2} (1998) 231. 
\bibitem{KS} S. Kachru \& E. Silverstein, Phys. Rev. Lett. {\bf 80} (1998)
4855. 
\bibitem{LNV} A. Lawrence, N. Nekrasov \& C. Vafa, Nucl. Phys. {\bf B533} 
(1998) 199. 
\bibitem{FV} P.H. Frampton \& C. Vafa, hep-th/9903226. 
\bibitem{F1} P.H. Frampton, Phys. Rev. {\bf D60} (1999) 085004.
\bibitem{FS} P.H. Frampton \& W.F. Shively, Phys. Lett. {\bf B454} (1999) 49.
\bibitem{F3} P.H. Frampton, hep-th/9908167. 
\bibitem{F2} P.H. Frampton, Phys. Rev. {\bf D60} (1999) 121901. 
\bibitem{FK} P.H. Frampton \& T.W. Kephart, Phys. Lett. {\bf B485} (2000) 403.
\bibitem{CST} C. Cs\'{a}ki, W. Skiba \& J. Terning, Phys. Rev. {\bf D61} 
(2000) 025019. 
\bibitem{mincer} S.G. Gorishny, S.A. Larin, L.R. Surguladze \& F.K. Tkachov, 
Comput. Phys. Commun. {\bf 55} (1989) 381. 
\bibitem{form} J.A.M. Vermaseren, ``{\sc FORM}'' version $2.3$, (CAN
Amsterdam, 1992).  
\bibitem{mincerform}
S.A. Larin, F.V. Tkachov \& J.A.M. Vermaseren, ``The Form version of
Mincer'', NIKHEF-H-91-18.
\bibitem{bfm} B.S. DeWitt, Phys. Rev. {\bf 162} (1967) 1195. 
\bibitem{tHbfm} G. 't Hooft, Acta Universitatis Wratislaviensis {\bf 368} 
(1976) 345, Proceedings of the 1975 Winter School of Theoretical Physics
held in Karpacz. 
\bibitem{DWbfm} B.S. DeWitt, in Proceedings of Quantum Gravity II, eds C. 
Isham, R. Penrose \& S. Sciama, (Oxford, 1980) 449.  
\bibitem{Bbfm} D.G. Boulware, Phys. Rev. {\bf D23} (1981) 389.  
\bibitem{Abbott:1981hw}
L.F. Abbott, Nucl. Phys. {\bf B185} (1981) 189.
\bibitem{Capper:1982tf} D.M. Capper \& A. MacLean, Nucl. Phys. {\bf B203} 
(1982) 413.
\bibitem{Grassi} P.A. Grassi, Nucl. Phys. {\bf B560} (1999) 499.
\bibitem{mathem}
S. Wolfram, {\sc Mathematica}, (Addison-Wesley, Reading, Mass, 1992). 
\bibitem{qgraf} P. Nogueira, J. Comput. Phys. {\bf 105} (1993) 279.
\bibitem{CUR} T. Curtright, Phys. Rev. {\bf D21} (1980) 1543. 
\bibitem{susybetas} L.V. Avdeev \& O.V. Tarasov, Phys. Lett. {\bf B112} (1982) 
356.
\bibitem{susy4a} L.V. Avdeev, O.V. Tarasov \& A.A. Vladimirov, Phys. Lett. 
{\bf B96} (1980) 94.
\bibitem{susy4b} 
M. Grisaru, M. Rocek \& W. Siegel, \prl 45 (1980) 1063; 
\npb {\bf B183} (1981) 141\semi
W.E. Caswell \& D. Zanon, Phys. Lett. {\bf B100} (1981) 152; 
Nucl. Phys. {\bf B182} (1981) 125.
\end{thebibliography}
\end{document}